\documentstyle[prd,aps,epsfig]{revtex}

\newdimen\nude\newbox\chek
\def\slash#1{\setbox\chek=\hbox{$#1$}\nude=\wd\chek#1{\kern-\nude/}}

\begin{document}

\draft

\title{Angular Dependence of the Radiative Gluon Spectrum\\ 
  and the Energy Loss of Hard Jets in QCD Media}

\preprint{BI-TP 99/16,
Bicocca-FT-99-21,
CU-TP-943, 
LPT-Orsay 99-50}

\author{R.~Baier}
\address{
Fakult\"{a}t f\"{u}r Physik, Universit\"{a}t Bielefeld,
D-33501 Bielefeld, Germany}

\author{Yu.L.~Dokshitzer\footnote{
Permanent address: Petersburg Nuclear Physics Institute, Gatchina 188350, 
St.~Petersburg, Russia}}
\address{INFN, Sezione di Milano, 20133 Milan, Italy}

\author{A.H.~Mueller\footnote{Supported in part by the U.S. Department of 
Energy under Grant DE-FG02-94ER-40819}}
\address{Physics Department, Columbia University, New York, NY 10027, USA}

\author{D.~Schiff\footnote{Laboratoire associ\'e du Centre National de 
la Recherche Scientifique}}
\address{LPT, Universit\'e Paris-Sud, B\^atiment 210, F-91405 Orsay, France}

\maketitle

\begin{abstract}
The induced momentum spectrum of soft gluons radiated from a high energy 
quark propagating through a QCD medium is derived in the BDMPS formalism. 
A calorimetric measurement for the medium dependent energy lost by a jet
 with opening angle $\theta_{{\rm cone}}$ is proposed.The fraction of this
 energy loss with respect to the integrated one appears to be the relevant 
observable.It exhibits  a universal behaviour  in terms of the variable 
$\theta^2_{{\rm cone}} L^3 \hat q$ where $L$ is the size of the medium and 
$\hat q$ the transport coefficient. Phenomenological implications for the 
differences between cold and hot QCD matter are discussed.

\end{abstract}

\pacs{12.38.Bx, 12.38.Mh, 24.85.+p, 25.75.-q}

\section{Introduction}
The medium induced energy loss of a hard (quark or gluon) jet traversing
matter - hot or cold - has recently been the subject of intensive interest
\cite{Gyulassy,Wang,Baier1,Baier2,Baier3,Baier4,Zakharov1,Zakharov2,Doksh,Lokhtin1,Lokhtin2,Wiedemann,Zakharov3,Baier5}.
The radiative energy loss has been found to be independent of the jet 
energy, for large energies, and growing as $L^2$ where $L$ is the 
extent of the medium \cite{Baier2,Baier3}. 
The order of magnitude of this effect in hot matter
may be expected to be large enough - compared to the case of cold 
nuclear matter - to lead to an observable and remarkable signal of the 
production of deconfined matter. 

The natural observable to measure this energy loss would then be the 
transverse momentum spectrum of hard jets produced in heavy-ion 
collisions. Jet quenching which has been discussed recently in 
\cite{Wang2,Levai}
is the manifestation of energy loss as seen in the suppression and 
change of shape of the jet spectrum compared with hadronic data. 
One may also think of measuring single particle inclusive spectra. 
Comparisons of high $p_\bot$ particle spectra in $p+p, 
p+A$ and $A+B$ collisions at SPS energies have already been performed and 
discussed \cite{Wang2}. 
The results seem to be difficult to interpret in view of the obvious model 
dependence of the theoretical prediction. 
This model dependence is associated with the relatively low energy and 
$p_\bot$ ranges which are currently accessible \cite{Levai}.
The situation should be much more favorable at RHIC. 
Jet production will be intensively studied at LHC. 

In Section 2 of the present paper we study the angular distribution of radiated
gluons which are believed to be the main source of energy loss. This allows for
quantitative predictions of the energy lost outside the cone defining the 
jet. This problem has already been
 considered in \cite{Doksh,Lokhtin1,Lokhtin2},
where the expression of the ``characteristic'' angle for gluon 
emission has been worked out. We work along the same lines but we perform a 
more complete calculation of the angular distribution.  
We concentrate on the realistic case of a hard jet produced in the medium.
We make the implicit assumption that the lifetime of the hot medium is 
large enough to allow for a large number of scatterings of the jet 
as it traverses the medium. 

In Section 3 we calculate the integrated energy loss outside an angular cone
with fixed opening angle, $\Delta E (\theta_{{\rm cone}})$,
and derive the expression for the ratio $R (\theta_{{\rm cone}} ) = 
{\Delta E (\theta_{{\rm cone}})/\Delta E}$, where 
$\Delta E$ is the completely integrated loss.
For simplicity of expression we take a convention where $\Delta E$ is a 
positive quantity.
 The surprising 
result is that $R (\theta_{{\rm cone}})$ is a universal function of the 
variable $\theta^2_{{\rm cone}} \hat q L^3$, where $\hat q$ is the 
transport coefficient characteristic of the medium. 

In Section 4 we give estimates and in particular we show that the energy 
loss is more collimated in the case of a hot QCD medium as compared to nuclear
matter. As a consequence the characteristic angle for gluon emission in 
a hot medium is quite small of the order of $10^\circ$ for a temperature of 
250 MeV. It is, however, possible to choose large enough angles of the 
order of $30^\circ$ and still have an appreciable energy loss.
We also make a comparison with the corresponding energy loss outside a 
cone in the vacuum. 
All the calculations are done on the partonic level with no attempt to include 
fragmentation of partons into hadrons. Fragmentation effects,
as well as effects due to plasma expansion \cite{Baier5}, could be 
important and should be studied. 

Technical details are summarized in Appendices A - C. 

\section{Momentum spectrum of radiated gluons} \label{momentum spectrum}

In this section we discuss the induced momentum spectrum of soft gluon 
emission $(x\rightarrow 0)$ from a fast quark jet. We assume that the quark is 
produced {\it inside} the medium by a hard scattering at time $t=0$. 
From the production point it propagates over a length $L$ of QCD matter, 
carrying out many scatterings with the medium. We follow closely the 
derivation and notation given in BDMPS \cite{Baier2,Baier3}
(denoted by I and II) and BDMS \cite{Baier4}
which we denote by III. 

In the case under consideration the spectrum per unit length $z$ of the 
medium consists of two terms: one corresponding to an 
``on-shell'' quark, as if that quark were entering the medium, and one 
corresponding to a hard production 
vertex as described in III (cf. Eq.(31)):
\begin{equation}\label{eq:2.1}
\frac{\omega dI}{d\omega dz d^2 \underline{U}} = \frac{\omega dI}{d\omega
dz d^2 \underline{U}} \,\, \Bigg|_{{\rm on-shell}} + 
\frac{\omega dI}{d\omega dz d^2 \underline{U}}\,\, \Bigg|_{{\rm vertex}}.
\end{equation}
$\omega$ denotes the soft gluon energy and $\underline U$ the scaled transverse
gluon momentum $\underline{U} = \underline{k}/\mu$ with $\mu$ the 
appropriate scale of the QCD potential. 

As follows from Eq.(31) in III
the induced gluon spectrum is 
\begin{eqnarray}\label{eq:2.2}
\frac{\omega dI}{d\omega dz d^2 \underline{U}} & = & 
\frac{\alpha_s C_F}{\pi^2L} 2 \,Re \,\int^L_0 \, dt_2 \int d^2 \underline{Q}
 \left\{ 
\int^{t_2}_0 \, dt_1 \rho\sigma \frac{N_c}{2C_F} f (\underline{U} + 
\underline{Q} , t_2 - t_1 ) + f_h (\underline{U} + \underline{Q} , t_2) 
 \right\}  \nonumber \\
&\times& \rho\sigma \frac{N_c}{2 C_F} 2 \left[ \frac{\underline{U} + 
\underline{Q}}{(\underline{U} + \underline{Q})^2} - \frac{\underline{U}}
{\underline{U}^2}\right] V (\underline{Q} ) \,
{\cal F}_{fsi} \, 
\Bigg|^{\tilde\kappa =0}_
{\tilde\kappa},
\end{eqnarray}
where $f_h$ differs from $f$ in that the gluon emission time has been
evaluated at the time of the hard interaction, $t=0$.
After we have converted (\ref{eq:2.2}) to impact parameter space the explicit 
relationship between  $f_h$ and $f$ will be given.   
In the soft $\omega$ limit the contributions from the diagrams (Fig. 1a-c)
are included. In comparison with the diagrams shown in Fig.~4 of III
we note that the soft limit leads to a vanishing of the sum of diagrams 
Fig. 4d-4j.


\begin{figure}[ht]
\centering
\epsfig{file=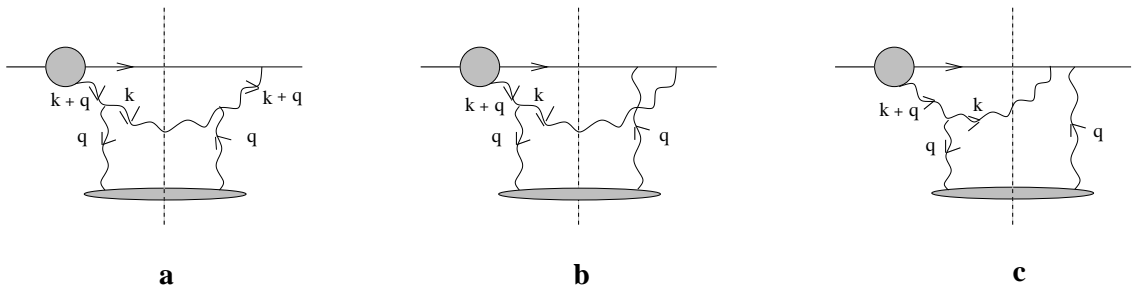,angle=-90, width=150mm}
\vskip .7cm
\caption{\label{fig:graph1} The diagrams representing the contributions
to the spectrum (\ref{eq:2.2}). }
\end{figure}

The quantities in (\ref{eq:2.2}), explained in III, have the following 
meaning: 
$\alpha_s C_F / \pi^2$ is the coupling of a gluon to a quark. The 
factor $\frac{N_c}{2C_F} f ( \underline{U} + \underline{Q}, t_2 - t_1)
\rho \sigma dt_1$ gives the number of scatterers in the medium, $\rho\sigma 
dt_1$, times the gluon emission amplitude at $t_1$, evolved to $t_2$. 
The momentum labels are shown in Fig.~1.
We describe the scattering in terms of the normalized cross section 
$V ( \underline{Q}) = \frac{1}{\pi\sigma} \frac{d\sigma}{d\underline{Q}^2}$ 
which depends on the scaled momentum $\underline{Q} = {\underline{q}}/{\mu}$. 
We note that the final state gluon carries a transverse 
momentum $\underline{k}$. 
The factor corresponding to the amplitude $f_0^\star$ in III
which sums the gluon emissions (Fig. 1a-c)
is $\int d^2 \underline{Q}\, V (\underline{Q}) \, 2 \left[ \frac{\underline{U} 
+ \underline{Q}}{(\underline{U}+\underline{Q})^2} - \frac{\underline{U}}
{\underline{U}^2}\right]$.
In order to eliminate the medium-independent factorisation contribution one 
has to perform a subtraction of the value of the integrals at 
$\tilde\kappa = 0$ \cite{Baier1,Baier2,Baier3,Baier4}.
The parameter $\tilde\kappa$ 
\begin{equation}\label{eq:2.3}
\tilde\kappa = \frac{2 C_F}{N_c} \frac{\lambda \mu^2}{2\omega},
\end{equation}
depends on medium properties, in particular
on the quark's mean free path $\lambda = 1/\rho\sigma$, as well as on the 
gluon energy.  

In dealing with (\ref{eq:2.2}) in contrast to the $\underline{U}$-integrated 
spectrum discussed in III
we now have to take into account the possibility that the 
emitted gluon may rescatter in the medium. For example the contribution shown 
in Fig. 1a
may now have final state interactions as illustrated in Fig. 2. 


\begin{figure}[ht]
\centering
\epsfig{file=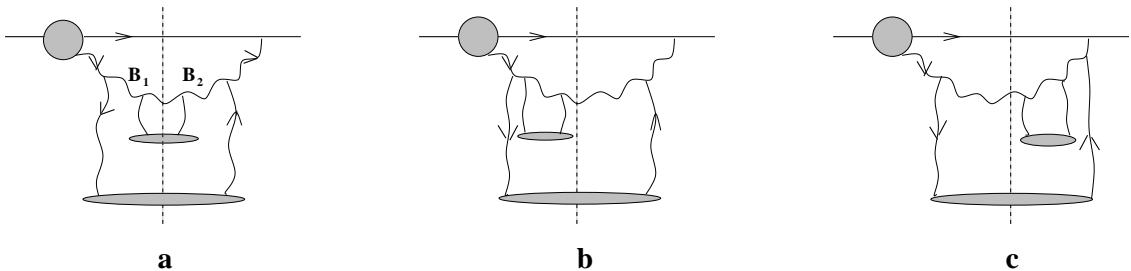,angle=-90, width=15cm}
\vskip .7cm
\caption{\label{fig:graph2}
Final state interactions contributing to Fig.~1a: (a) real, (b) and (c)
virtual interactions.
 }
\end{figure}

This is taken care of by the final state interaction amplitude 
${\cal F}_{fsi}$ 
in (\ref{eq:2.2}) which is most conveniently expressed in 
impact parameter space. One goes to impact parameter space by taking the 
Fourier transform, for example for the amplitude $f (\underline{U}, t)$, 
\begin{equation}\label{eq:2.4}
f ( \underline{U} , t) = \int \, \frac{d^2 \underline{B}}{(2\pi)^2} \,
e^{i \underline{B} \cdot \underline{U}} \tilde f (\underline{B}, t).
\end{equation}
As in III we rescale the time variable $t = \frac{2 C_F}{N_c} \lambda\tau$ and
define
\begin{equation} \label{eq:2.5}
\tau_0 = N_c L / 2 C_F \lambda.
\end{equation}
Following our analysis of parton $p_\bot$-broadening in II 
we write the final state interaction amplitude as 
\begin{equation}\label{eq:2.6}
{\cal F}_{fsi} ( \underline{B}_1 - \underline{B}_2 , \tau ) = \exp \, \left[
-\frac{1}{2} ( \underline{B}_1 - \underline{B}_2)^2 \tau\tilde v \right].
\end{equation}

In II the distribution in transverse momentum and the corresponding 
distribution in impact parameter were given for a single parton going through 
a medium. Here we have a quark-gluon system passing through the 
medium. However, interactions of the quark with the medium cancel between 
real and virtual interactions leaving only the interactions of the gluon
with the medium as in II. 
These interactions can be real interactions, as shown in Fig. 2a
where both the gluon in the amplitude at impact parameter $\underline{B}_1$ 
and in the complex conjugate amplitude at $\underline{B}_2$ interact with the
medium, or they can be virtual interactions as shown in Figs. 2b and 2c. 
The real interactions give a factor $\tilde V ( \underline{B}_1 - 
\underline{B}_2)$ while the virtual interactions give the absorption of the 
gluon wave and a factor of $- \tilde V (0) = -1$. 
$\tilde V (\underline{B})$ is the Fourier transform of the potential 
$V (\underline{Q})$.
These terms combine 
together giving at small $\underline{B}_1 - \underline{B}_2$  
\begin{equation}\label{eq:2.7}
1 - \tilde V (\underline{B}_1 - \underline{B}_2) \simeq \frac{1}{4}
(\underline{B}_1 - \underline{B}_2)^2 \, \tilde v.
\end{equation}
$\tilde v$ has a logarithmic dependence on $\underline{B}^2$.
In the following we neglect the variation of the logarithm, an approximation
which is good except when large transverse momentum rescatterings occur
(see II). 
The factor in (\ref{eq:2.7}) then exponentiates, as in II, to give 
(\ref{eq:2.6}), where the $\frac{1}{2}$ in the exponent (\ref{eq:2.6}) 
differs from the $\frac{1}{4}$ in Eq.(\ref{eq:3.3}) of II because of 
the difference of $\tau$ from $t$ used there. 
Using (\ref{eq:2.4}) and (\ref{eq:2.6}) in (\ref{eq:2.2}) we obtain the 
spectrum in terms of impact parameter integrals, 
\begin{eqnarray}\label{eq:2.8}
\frac{\omega d I}{d\omega dz d^2 \underline{U}} & = & 
\frac{\alpha_s C_F}{\pi^2 L} 2 Re \, \int^{\tau_0}_0 \, d \tau_2 \, \int \,
\frac{d^2 B_1}{(2\pi)^2} \frac{d^2 B_2}{(2\pi)^2} \, 
e^{i(\underline{B}_1 - \underline{B}_2) \cdot \underline{U}} \,
\left\{ \int^{\tau_2}_0 \, d \tau_1 \tilde f (\underline{B}_1 , \tau_2 - 
\tau_1) + \frac{2}{\tilde v \underline{B}_1^2} \tilde f (\underline{B}_1 , 
\tau_2 ) \right\}  \nonumber \\
& \times & \frac{4 \pi i \underline{B}_2}{\underline{B}_2^2} \left[ \tilde V (
\underline{B}_1 - \underline{B}_2) - \tilde V ( \underline{B}_1) \right]
e^{- \frac{\tilde v}{2} (\underline{B}_1 - \underline{B}_2)^2 (\tau_0 - 
\tau_2)} \Bigg|^{\tilde\kappa = 0}_{\tilde\kappa} .
\end{eqnarray}
We have used $\tilde f_h ( \underline{B}, \tau) = \frac{2}{\tilde v 
\underline{B}^2} 
\tilde f ( \underline{B}, \tau)$ which follows from the fact that $\tilde f_h$
and $\tilde f$ have the same time evolution while 
the initial conditions satisfy
$\tilde f_h (
\underline{B} , 0) = \frac{2}{\tilde v \underline{B}^2} \tilde f 
(\underline{B},0)$ as 
can be seen by comparing Eqs.(16) and (17b)
of III in the soft gluon limit. 

When integrating (\ref{eq:2.8}) with respect to the gluon momentum 
$\underline{U}$ it is straightforward to reproduce the energy spectrum given 
e.g. in Eq.(34) of III.
Next we recall that the amplitude $\tilde f (\underline{B} , \tau)$ is given 
by (cf. Eq.(38) in III)
\begin{equation}\label{eq:2.9}
\tilde f (\underline{B}, \tau) = - \frac{i \pi \tilde v}{\cos^2 \omega_0 
\tau} \underline{B} \exp \left( - \frac{i}{2} m \omega_0 \underline{B}^2 
\tan \omega_0 \tau \right), 
\end{equation} 
with 
\begin{equation}\label{eq:2.10}
m = - 1 / 2 \tilde\kappa \,\,\, {\rm and}\,\,\, \omega_0 = 
\sqrt{2i \tilde\kappa \tilde v}, 
\end{equation}
together with $\tilde\kappa$ given in (\ref{eq:2.3}). 
In this approximation (leading logarithm and small $x$) after 
substituting (\ref{eq:2.9}) into (\ref{eq:2.8}) we find a convenient 
expression 
for the momentum spectrum per unit length
\begin{equation}\label{eq:2.11}
\frac{\omega dI}{d\omega dz d^2 \underline{U}} = \frac{2 \alpha_s C_F}{L} 
\tilde v^2 Re\,
\int^{\tau_0}_0 \, \frac{d\tau}{\cos^2 \omega_0 \tau} \, \left\{
\int^{\tau_0 - \tau}_0 \, d\tau_1 \, I (\underline{U}, \alpha, \beta) + 
\int^{\tau_0 - \tau}_{-\infty} \, d\tau_1 \, I (\underline{U} ,
 \alpha + \beta, 
\beta^\prime) \right\} \Bigg|^{\tilde\kappa = 0}_{\tilde\kappa} , 
\end{equation}  
with the time dependent variables
\begin{eqnarray}\label{eq:2.12}
\alpha & = & \frac{i}{2} m \omega_0 \tan \omega_0 \tau = \frac{\tilde v}
{2} \frac{\tan\omega_0 \tau}{\omega_0} , \nonumber \\
\beta & = & \frac{\tilde v}{2} ( \tau_0 - \tau - \tau_1 ),  \\
\beta^\prime & = & \frac{\tilde v}{2} ( \tau_0 - \tau) . \nonumber
\end{eqnarray} 
After using (\ref{eq:2.7}) in (\ref{eq:2.8}) the $\underbar{B}$-space 
integrals become
\begin{eqnarray}\label{eq:2.13}
I ( \underline{U}, \alpha, \beta) &=& 
\int \, \frac{d^2 \underline{B}_1}{(2\pi)^2} \, \frac{d^2 \underline{B}_2}
{(2\pi)^2} \, e^{i ( \underline{B}_1 - \underline{B}_2 ) \cdot \underline{U}}
\nonumber \\
&\times & \frac{\underline{B}_2 \cdot \underline{B}_1}{\underline{B}_2^2}
\left[ \underline{B}_1^2 - ( \underline{B}_1 - \underline{B}_2 )^2 \right]
\exp \left[ - \alpha \underline{B}_1^2 - \beta (\underline{B}_1 - 
\underline{B}_2 )^2 \right] .
\end{eqnarray}
It is evaluated in Appendix A with the explicit result (\ref{eq:A.8}),
\begin{equation}\label{eq:2.14}
I ( \underline{U} , \alpha , \beta ) = \frac{1}{16\pi}\, \frac{1}{\alpha^2 
(\alpha + \beta )} e^{-\frac{\underline{U}^2}{4 (\alpha + \beta)}}. 
\end{equation}
From the induced momentum spectrum (\ref{eq:2.11}), (\ref{eq:2.12})
 one may rederive the $\omega$-spectrum per unit length 
$\omega dI / d\omega dz$ for soft gluon emission off quark jets 
produced inside the medium. After performing the exponential 
$\underline{U}$-integration and the $\tau, \tau_1$ integrals explicitly the 
result of III is reproduced, namely 
\begin{equation}\label{eq:2.15}
\frac{\omega d I}{d\omega dz} = \frac{2 \alpha_s C_F}{\pi L} \, \ln 
|\cos \omega_0 \tau_0 | . 
\end{equation}
We note from (\ref{eq:2.5}) and (\ref{eq:2.10}) that $\omega_0 \tau_0$ is 
expressed in terms of medium-dependent quantities and the gluon 
energy $\omega$ by 
\begin{eqnarray}\label{eq:2.16}
(\omega_0 \tau_0 )^2 &=& i \frac{N_c}{2C_F} \, \frac{\mu^2 \tilde v}{\lambda}
\, L^2 / \omega \nonumber \\
&=& i \frac{N_c}{2 C_F} \, \hat q L^2 / \omega , 
\end{eqnarray}
where we introduce the transport coefficient \cite{Baier2,Baier3}
\begin{equation}\label{eq:2.17}
\hat q \equiv \frac{\mu^2\tilde v}{\lambda} = \rho \, \int^{1/\underline{B}^2}
_0 \, d \underline{Q}^2 \, \underline{Q}^2 \, \frac{d\sigma}{d\underline{Q}^2}.
\end{equation}
Integrating (\ref{eq:2.15}) over $\omega$ the energy loss per unit length is 
obtained as
\begin{equation}\label{eq:2.18}
- dE/dz = \int^\infty_0 \, \frac{\omega dI}{d\omega dz} \, 
d\omega = \frac{\alpha_s N_C}{4} \, \hat q L ,
\end{equation}
so long as $E > E_{cr} \simeq \hat q L^2$ \cite{Baier2}.

\section{Induced radiative energy loss of a hard quark jet in a finite cone}


Let us consider a typical calorimetric measurement. We have in mind a hard
quark jet of high energy $E$ produced by a hard 
scattering in a dense QCD medium and propagating through it over a 
distance $L$. 
The quark loses energy
by gluon radiation induced by multiple scattering. 
In the following we calculate the integrated loss 
{\it outside} an angular cone of opening angle $\theta_{{\rm cone}}$
(Fig. 3),
\begin{equation}\label{eq:3.1}
\Delta E (\theta_{{\rm cone}}) = L \, \int^\infty_0 \, d\omega\,
\int^\pi_{\theta_{{\rm cone}}} \, \frac{\omega dI}{d\omega dz d\theta} 
d\theta .
\end{equation}


\begin{figure}[ht]
\centering
\epsfig{file=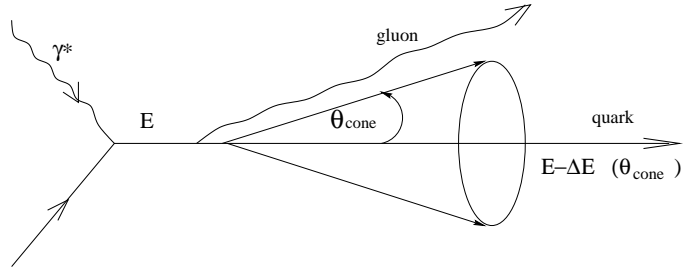,angle=-90,width=9cm}
\vskip  .7cm
\caption{\label{fig:graph3} 
Hard process producing a quark jet. The gluon is emitted outside
the cone with angle $\theta_{{\rm cone}}$.}
\end{figure}

\noindent
We note that for $\theta_{{\rm cone}} = 0$ the total loss (\ref{eq:2.18})
is obtained, namely \cite{Baier4}
\begin{equation}\label{eq:3.2}
\Delta E = \frac{\alpha_s N_c}{4} \hat q L^2 . 
\end{equation}
In detail we consider the normalized loss by defining the ratio 
\begin{equation}\label{eq:3.3}
R (\theta_{{\rm cone}} ) = \frac{\Delta E (\theta_{{\rm cone}})}
{\Delta E} , 
\end{equation}
which may be decomposed into 
\begin{equation}\label{eq:3.4}
R (\theta_{{\rm cone}}) = R_{{\rm on-shell}} (\theta_{{\rm cone}}) + 
R_{{\rm vertex}} (\theta_{{\rm cone}} )
\end{equation}
according to the treatment in the previous section. 

\noindent
Defining ${\underline{U}^2_{{\rm cone}}}$
in the approximation of small cone angles
\begin{equation}\label{eq:3.5a}
\underline{U}^2_{\rm cone} \simeq \frac{\omega^2}{\mu^2} \theta^2_{\rm cone},
\end{equation}
the $\theta$ integral in (\ref{eq:3.1}) may be performed using
\begin{equation}\label{eq:3.5}
\int^\infty_{\underline{U}^2_{{\rm cone}}} \, d^2 \underline{U} \, I 
(\underline{U} , 
\alpha , \beta ) = \frac{1}{4\pi \alpha^2}
 \exp \left[{-\frac{\omega^2 \theta^2_{{\rm 
cone}}}{4 \mu^2 (\alpha + \beta )}} \right] . 
\end{equation}
The remaining $\omega , \tau$ and $\tau_1$ integrations cannot be done 
analytically. It is convenient to change variables by 
introducing dimensionless ones:
\begin{eqnarray}\label{eq:3.6}
\tau &\equiv & \tau_0 y , \nonumber \\
\tau_1 &\equiv & \tau_0 z , \\
\omega_0 \tau &\equiv & (1 + i ) x, \nonumber
\end{eqnarray}
which implies that the gluon energy is expressed by 
\begin{equation}\label{eq:3.7}
\omega = \frac{C_F}{N_c} \lambda \mu^2 \tilde v \tau^2_0 (y / x)^2 .
\end{equation}
As a nice consequence, the ratio $R (\theta_{{\rm cone}})$
turns out to depend  on one single dimensionless variable
\begin{equation}\label{eq:3.8}
R = R (c (L) \theta_{{\rm cone}} ), 
\end{equation}
where
\begin{equation}\label{eq:3.9}
c^2 (L) = \frac{N_c}{2C_F} \hat q \left( L / 2 \right)^3 .
\end{equation}
The ``scaling behaviour'' of $R$
means that the medium and size dependence is universally contained in the
function $c (L)$, which is a function of the transport coefficient $\hat q$ 
(\ref{eq:2.17}) and of the length $L$, as defined by (\ref{eq:3.9}). 
For fixed $L$ the medium properties are exclusively described by $\hat q$ 
which is different for cold and hot matter (cf. our 
discussion in Sect.~5 of II).
The fact that $\theta_{{\rm cone}}$ scales as $1/c(L)$ may be understood
from the following physical argument \cite{Doksh}:
the radiative energy loss of a quark jet is dominated by gluons having 
$\omega \simeq \hat q L^2$ 
(cf. Eq.(6.9) in I).
 The angle that the emitted gluon makes with the quark is $\theta = 
{\underline{k}}/{\omega}$
 with $\underline{k}$ the gluon transverse momentum. But 
$\underline{k}^2 \simeq \hat q L$ so that the typical gluon angle will be 
$\theta^2
\simeq \frac{1}{\hat q L^3}$ or $\theta^2 \simeq \left[ \frac{(\hat q)^{
1/3}}{\omega}\right]^{3/2}$.

\begin{figure}
\centering
\epsfig{bbllx=35,bblly=210,bburx=515,bbury=625,
file=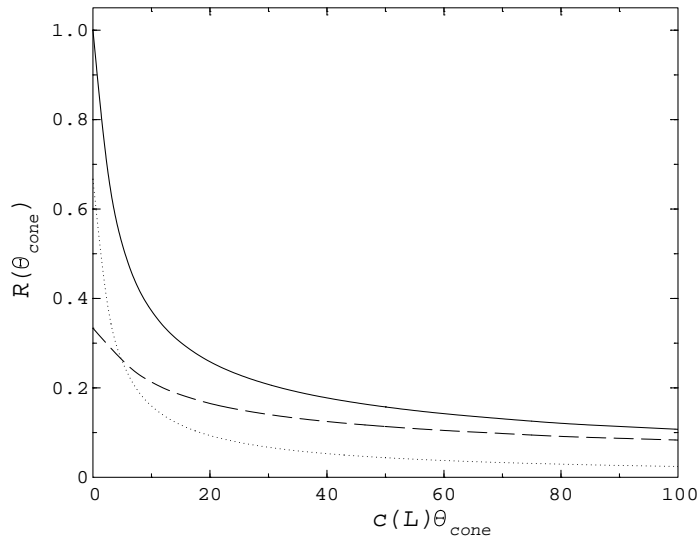,width=96mm,height=80mm}
\caption{\label{fig:graph4}Fractional induced loss $R(\theta_{{\rm cone}})$
(solid curve) as a function of $ c(L)\theta_{{\rm cone}}$, and decomposed into 
$R_{{\rm on-shell}}$ (dashed curve) and $R_{{\rm vertex}}$ (dotted curve). }
\end{figure}

For completeness the explicit dependence of $R$ on $c(L) \theta_{{\rm cone}}$
is given in Appendix B.
The numerical evaluation of this dependence is shown 
in Fig.~4 \cite{Mathem}.
The following remarks may be useful. Obviously by definition $R (\theta_{{\rm
cone}} = 0 ) = 1$, where we know from III that 
\begin{equation}\label{eq:3.11}
R_{{\rm on-shell}}(\theta_
{{\rm cone}}=0 ) = 1/3 \,\,\,\,\,\, {\rm and}\,\,\,\,\,\,R_{{\rm vertex}} 
(\theta_{{\rm cone}}=0) =2/3  . 
\end{equation}
At large angles, $c(L) \theta_{{\rm cone}} \gg 1$,
we find (see Appendix C)
the power behaviour
\begin{equation}\label{eq:3.12}
R (\theta_{{\rm cone}}) \longrightarrow \frac{4 \Gamma (1/4)}{5\pi} 
\frac{1}{(c( L)\theta_{{\rm cone}})^{1/2}} , 
\end{equation}
coming exclusively from $R_{{\rm on-shell}}$. $R_{{\rm vertex}} (\theta_{
{\rm cone}})$ vanishes faster than the dependence given in 
(\ref{eq:3.12}), as may be seen from (\ref{eq:C.6}).

The ratio $R(\theta_{{\rm cone}})$ is also universal in the sense
that it is the same for an energetic quark as well as for a gluon jet,
since the function $c(L)$ depends on the product $C_F \lambda =
N_c \lambda_{gluon}  $( $ = C_R \lambda_R$ for colour representation R).
It is the integrated loss, however, which is bigger by the factor
$N_c/C_F$ for the gluon than for the quark jet.

The ``narrowness'' of the angular distribution can be read off from Fig. 4:
e.g. $R (\theta_{{\rm cone}})$ decreases from $1$ to $0.4$, when $c
(L)\theta_{{\rm cone}}$ is increased from zero to $10$. Explicit values for
$c(L)$ are discussed in Sect. 4 when comparing hot and cold QCD matter
for fixed $L$
\begin{equation}\label{eq:3.13}
c (L) \Big|_{{\rm HOT}} \gg c(L) \big|_{{\rm COLD}}, 
\end{equation}
so that $R(\theta_{{\rm cone}})
\Big|_{{\rm HOT}}$ is much narrower than $R(\theta_{{\rm cone}})\Big|_
{{\rm COLD}}$ when considered as a function of $\theta_{{\rm cone}}$.


\section{Phenomenology}


We know from I - III that the transport coefficient $\hat q$ 
(\ref{eq:2.17})
controls the medium induced energy loss (\ref{eq:3.2}) as well as the 
$p_\bot$-broadening of high energy partons:
both quantities are directly proportional to $\hat q$. The medium dependence 
of the distribution $R(\theta_{{\rm cone}})$ in (\ref{eq:3.8})
(see also Appendix B) 
is determined by the coefficient $c(L)$ which according to (\ref{eq:3.9})
depends on the square root of $\hat q$.  
In this section we estimate 
the magnitude of $c(L)$, with the aim of finding differences in the 
angular distributions $R(\theta_{{\rm cone}})$ for cold and hot 
matter. In the following we will confirm the statement given in 
(\ref{eq:3.13}) quantitatively.  
In order to be able to compare with our discussion already given in II 
(and in III) 
we keep the same numerical values for the quantity $\hat q$ as used there. 

(i) For hot matter at reference temperature $T = 250$ MeV we take
\begin{equation}\label{eq:5.1}
\hat q \simeq 0.1 \, {\rm GeV}^3 .
\end{equation}
This value is deduced from (\ref{eq:2.17}) using the screened one-gluon 
exchange cross section,
e.g. for two flavours one derives \cite{Wang}
\begin{equation}\label{eq:5.2}
\hat q \simeq \frac{48}{\pi} \zeta (3) \tilde v \, \alpha^2_s \, T^3, \,\,
\zeta (3) \simeq 1.202. 
\end{equation}
We use a typical value for the dimensionless quantity $\tilde v \simeq 2$. 
Eq.(\ref{eq:5.2}) shows explicitly the temperature dependence of 
$\hat q$. For the values of $T$ under consideration (\ref{eq:5.1}) leads to 
a rather large value of $c(L)$, namely
\begin{equation}\label{eq:5.3}
c(L) \Big|_{{\rm HOT}} \simeq 40 \left( L / 10 fm \right)^{3/2} , 
\end{equation}
where we use $\alpha_s \simeq 1/3$ as in II. 

This large value of $c(L)$ implies that for cone angles $\theta_{{\rm cone}} 
\geq 30^\circ$ the normalized distribution $R(\theta_{{\rm cone}})$ is well
approximated by its asymptotic form given by (\ref{eq:3.12}).
For large $c(L)\theta_{{\rm cone}}$ the fraction of the induced energy loss  
$R(c(L)\theta_{{\rm cone}})$ decreases as $(TL)^{-3/4}$ for 
increasing temperature $T$ and length $L$ (for fixed $\theta_{{\rm cone}}$).
We also note that the energy loss outside the cone quickly drops with 
increasing $\theta_{{\rm cone}}$ (cf. Figs. 4 and 5):
for $\theta_{{\rm cone}}\simeq 10^\circ \, (40^\circ)$ the fraction $R$ is 
reduced to $40\% \, (20\%)$ of the total loss, for which we find a value of 
\begin{equation}\label{eq:5.4}
\Delta E\simeq 60 \, {\rm GeV} \, \left( L / 10 fm \right)^2
\end{equation}
for a quark jet, according to Eq.(\ref{eq:3.2}).


\begin{figure}
\centering
\epsfig{bbllx=45,bblly=205,bburx=510,bbury=610,
file=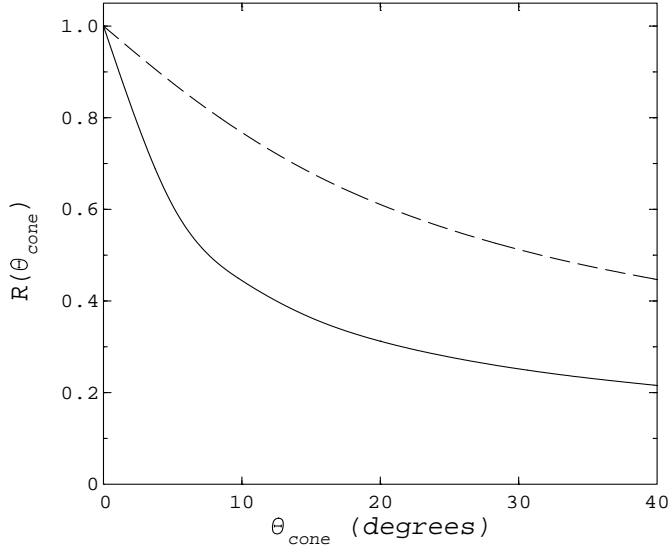,width=96mm,height=80mm}
\caption{\label{fig:graph5}Medium induced (normalized) energy loss 
distribution as a function of cone angle $\theta_{{\rm cone}}$ for hot 
($T=250$ MeV) (solid curve) and cold matter (dashed curve) at fixed 
length $L=10 fm$.}
\end{figure}

\noindent
Indeed the medium dependent distribution $R(\theta_{{\rm cone}})$ is rather 
narrow on the partonic level (Fig. 5).

(ii) For cold matter we quote from II the typical value
\begin{eqnarray}\label{eq:5.5}
\hat q & \simeq & \frac{1}{25} \alpha_s [x G (x) ] \,{\rm GeV}^2/fm 
\nonumber \\
&\simeq & 0.005 \, {\rm GeV}^3 , 
\end{eqnarray}
with a gluon distribution, $x G (x) \simeq 1 - 2$. Compared with 
(\ref{eq:5.4}) we find 
\begin{equation}\label{eq:5.6}
 \Delta E \simeq 3.5 \, {\rm GeV} \, \left( L / 10 fm \right)^2 ,
\end{equation}
and accordingly a much smaller value for 
\begin{equation}\label{eq:5.7}
c (L) \big|_{{\rm COLD}} \simeq 10 \, \left( L / 10 fm \right)^{3/2} .
\end{equation}
This confirms Eq.(\ref{eq:3.13}) quantitatively, namely 
\begin{equation}\label{eq:5.8}
c (L) \Big|_{{\rm HOT}} \simeq 4 \, c (L) \Big|_{{\rm COLD}}
\end{equation}
at $T = 250$ MeV. 
When comparing the energy loss in hot and cold matter we observe from 
(\ref{eq:5.4}) and (\ref{eq:5.6}) that the induced loss in cold matter is much 
smaller (by a factor of about 15) than in hot matter, and that $R (\theta_
{{\rm cone}}) \Big|_{{\rm COLD}}$ is broader than $R(\theta_{{\rm cone}})
\Big|_{{\rm HOT}}$, which can be seen from Fig. 5. 

As a conclusion we expect that the medium-induced $\Delta E
 (\theta_{{\rm cone}})$ for energetic jets can be large  in 
{\it hot QCD} matter and although collimated, still appreciably larger than in
cold matter even for cone sizes of order
 $\theta_{{\rm cone}} \simeq 30^\circ$.

So far we have discussed the {\it medium-induced} energy loss and its angular 
distribution. Concerning the total energy loss of a jet of a given
cone size it is 
important to take into account the {\it medium independent} part. It 
corresponds to the $\tilde\kappa = 0$ contribution, which we have 
subtracted in Eq.(\ref{eq:2.2}).
For the case of a fast quark produced in the medium the $\tilde\kappa = 0$
soft gluon radiation spectrum is given in leading order by the well known 
Born term, 
\begin{equation}\label{eq:5.9}
\frac{\omega d I^{\tilde\kappa = 0}}{d\omega dz d^2 \underline{U}}
= \frac{\alpha_s C_F}{\pi^2 L} \frac{1}{\underline{U}^2} ,
\end{equation}
valid for single gluon emission off a quark in the vacuum, and therefore
truly medium independent. This result may be obtained as follows: the 
$\tilde\kappa = 0$ factorization term can be realized as the 
emission of large $\omega$ gluons.
These gluons are produced with large formation times, such that 
the dominant gluon emission takes place {\it after} the energetic quark has
propagated through the medium, namely far outside the medium. A more careful 
analysis confirms the result (\ref{eq:5.9}).
The angular dependence of the corresponding energy loss of a quark jet in a 
cone may be estimated from (\ref{eq:5.9}), 
\begin{eqnarray}\label{eq:5.10}
 \Delta E^{\tilde\kappa = 0} ( \theta_{{\rm cone}}) &=&
L \, \int^E_0 \, d\omega \, \int^{\underline{U}^2_{{\rm max}}}_
{\underline{U}^2_{{\rm cone}}} \,
\frac{\omega d I^{\tilde\kappa = 0}}{ d\omega dz d^2 
\underline{U}} \, d^2 \underline{U} \nonumber \\
&\simeq& 2 \frac{\alpha_s C
_F}{\pi} E \, \ln \left( \frac{\theta_{{\rm max}}}{\theta_{{\rm cone}}}\right)
, 
\end{eqnarray}
using a constant $\alpha_s$. Since in this case the loss is sensitive to gluon 
energies $\omega = {\cal O} (E)$, one should take into account the 
quark $\rightarrow$ gluon splitting function $\frac{x}{2} \frac{1 + (1 - x)^2}
{x}$, which amounts to replacing  $E$ by $2/3 E$ in 
(\ref{eq:5.10}). 

Keeping in mind that the following estimates are 
based on the leading logarithm approximation, we summarize in Fig. 6
the medium-induced (for a hot medium with $T = 250$ MeV) and the medium 
independent energy losses.


\begin{figure}
\centering
\epsfig{bbllx=30,bblly=205,bburx=495,bbury=600,
file=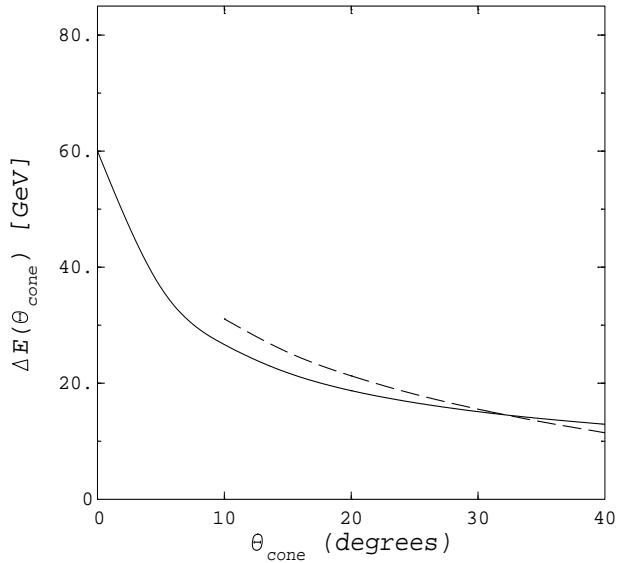,width=88mm,height=80mm}
\caption{\label{fig:graph6} Energy loss in a hot medium, $T=250 MeV$,
as a function of $\theta_{{\rm cone}}$.
Dashed curve represents (\ref{eq:5.10}) for $E=250 GeV$. $L=10 fm $.  }
\end{figure}

For the strict validity of (\ref{eq:3.1}) we recall the condition \cite{Baier2}
that $E > E_{{\rm crit}} \simeq \hat q L^2$, which requires very large 
values of the quark energy: $E > 250 \left( L / 10 fm \right)^2$ GeV, 
when $T\simeq 250$ MeV. Such  
energetic jets may indeed be produced in hard scatterings
 in heavy-ion collisions at the LHC at CERN. 
For $L = 10 fm$ this jet energy leads to $\Delta E^{\tilde\kappa = 0}
(\theta_{{\rm cone}}) \simeq 15 \ln \left( \frac{\theta_{{\rm max}}}{\theta_
{{\rm cone}}} \right)$ GeV. 
To get this estimate we have taken 
$\alpha_s \simeq 1/10$ due to the large characteristic
 transverse momentum of the emitted gluon.
The $\theta_{{\rm cone}}$ dependence is shown in Fig. 6
for the two contributions: over the limited range of $10^\circ\leq
\theta_{{\rm cone}}
\leq 40^\circ$ they turn out to be not very different $(\theta_{{\rm max}}$ is 
taken ${\cal O} (\pi/2))$.
It is, however, important to keep in mind that (\ref{eq:5.10}) and 
(\ref{eq:3.2}) differ with respect to their dependence on $L$. From
 (\ref{eq:5.10})
$\Delta E^{\tilde\kappa = 0} (\theta_{{\rm cone}})$ does not depend
on $L$ at all. The $L$ dependence of  
$\Delta E (\theta_{{\rm cone}}) = R (\theta_{{\rm cone}}) \Delta E$ 
comes from two sources. 
It is proportional to $L^2$ from $\Delta E$,
with an extra $L$ dependence contained in 
$R(\theta_{{\rm cone}})$ at fixed cone angle due to
$c(L)$, Eq.({\ref{eq:3.9}).

\vspace{0.5cm} 
\subsection*{Acknowledgments} 
\noindent 
This research is supported in part 
by Deutsche Forschungsgemeinschaft (DFG), Contract Ka 1198/4-1.
We thank Marcus Dirks for his help with the numerical analysis.


\begin{appendix}
\section{$\underline{B}$-space integration} \label{appendix-a}

The gluon momentum spectrum is expressed in terms of an integral $I 
(\underline{U} , \alpha , \beta )$ (Eq.(\ref{eq:2.13})). It can be 
reduced to 
\begin{equation}\label{eq:A.1}
I ( \underline{U}, \alpha , \beta ) = \left( \frac{\partial}{\partial\beta} 
- \frac{\partial}{\partial\alpha} \right) J ( \underline{U} , \alpha , 
\beta ), 
\end{equation}
where
\begin{eqnarray}\label{eq:A.2}
J ( \underline{U} , \alpha , \beta )&=& \int \, \frac{d^2 B_1}{(2\pi)^2} \,
\frac{d^2 B_2}{(2\pi)^2} \, e^{i (\underline{B}_1 - \underline{B}_2)\cdot
\underline{U}} \nonumber \\
&\times& \underline{B}_1 \cdot \frac{\underline{B}_2}{\underline{B}_2^2} \, 
\exp \, \left[ - \alpha \underline{B}_1^2 - \beta ( \underline{B}_1 - 
\underline{B}_2 )^2 \right] . 
\end{eqnarray}
In order to perform the $\underline{B}_1 , \underline{B}_2$ 
integrations we note that 
\begin{equation}\label{eq:A.3}
\frac{\underline{B}_2}{\underline{B}_2^2} = \frac{1}{2} \vec\nabla_{B_2} \, 
\ln \, \underline{B}_2^2 , 
\end{equation}
and 
\begin{equation}\label{eq:A.4}
\underline{B}_1 \, \exp \left[ - \alpha \underline{B}_1^2 - \beta (
\underline{B}_1 - \underline{B}_2 )^2 \right] = - \frac{1}{2\alpha} 
\left( \vec\nabla_{B_1} + \vec\nabla_{B_2} \right) \, \exp \, \left[
-\alpha \underline{B}_1^2 - \beta ( \underline{B}_1 - \underline{B}_2)^2 
\right] . 
\end{equation}
These identities allow us to write 
\begin{eqnarray}\label{eq:A.5}
J ( \underline{U} , \alpha , \beta ) &= &
- \frac{1}{4\alpha} \, \int \, \frac{d^2 \underline{B}_1}{(2\pi)^2} \,
\frac{d^2 \underline{B}_2}{(2\pi)^2} \, e^{i ( \underline{B
}_1 - \underline{B}_2)\cdot \underline{U}} \nonumber \\
&\times & \left( \vec\nabla_{B_2} \, \ln \, \underline{B}_2^2 \right) 
\left( \vec\nabla_{B_1} + \vec\nabla_{B_2} \right) \, \exp \, \left[ - 
\alpha \underline{B}_1^2 - \beta ( \underline{B}_1 - \underline{B}_2 )^2
\right] . 
\end{eqnarray}
After partial integration and using 
\begin{eqnarray}\label{eq:A.6}
& & \left( \vec\nabla_{B_1} + \vec\nabla_{B_2} \right) \, e^{i (\underline{B}_1
-\underline{B}_2 ) \cdot \underline{U}} = 0 , \nonumber \\
& & \vec\nabla_{B_2}^2 \, \ln \, \underline{B}_2^2 = 4 \pi \delta^2 (
\underline{B}_2), 
\end{eqnarray}
one finds 
\begin{eqnarray}\label{eq:A.7}
J ( \underline{U} , \alpha , \beta ) &= &
\frac{1}{16 \pi^3 \alpha} \, \int \, d^2 \underline{B}\, e^{i \underline{B} 
\cdot \underline{U}} \, e^{-(\alpha + \beta ) \underline{B}^2} \nonumber \\
& & \nonumber \\
& = & \frac{1}{16\pi^2} \, \frac{1}{\alpha (\alpha + \beta)} \, 
\exp \left[{- \frac{\underline{U}^2}{4(\alpha + \beta)}} \right] . 
\end{eqnarray}
Finally the function $I ( \underline{U} , \alpha , \beta ) $ becomes 
\begin{equation}\label{eq:A.8}
I ( \underline{U} , \alpha , \beta ) = \frac{1}{16\pi^2} \, \frac{1}{\alpha^2
(\alpha + \beta )} \,
 \exp \left[{-\frac{\underline{U}^2}{4(\alpha + \beta ) }} \right] \,\, .
\end{equation}

\section{Explicit expression for $R(\theta_{{\rm cone}})$} \label{appendix-b}


Here we summarize the explicit expressions for 
$R_{{\rm on-shell}} (\theta_{{\rm cone}})$ and  
  $R_{{\rm vertex}} (\theta_{{\rm cone}})$.
They are obtained  from Eq.(\ref{eq:2.11}) performing the integrations
defined in (\ref{eq:3.1}) by applying the change of variables (\ref{eq:3.6}).

The results are:
\begin{eqnarray}\label{eq:B.1}
R_{{\rm on-shell}} (\theta_{{\rm cone}}) &=&
\frac{4}{\pi} \, \int^\infty_0 \, \frac{dx}{x^3} \, \int^1_0 \, dy \, 
\int^{1-y}_0 \, dz \left\{ \exp \left[ - \frac{c^2(L)\theta^2_{{\rm cone}}}
{(1-z)} \, \left( \frac{y}{x}\right)^4\right] \right. \nonumber \\
&-& Re \left. \left[ \frac{\left( (1+i)x\right)^2}{\sin^2 (1+i)x}\, \exp \left[
- \frac{c^2 (L) \theta^2_{{\rm cone}}}{y \frac{\tan (1+i)x}{(1+i)x} + 1 - z 
-y} \, \left( \frac{y}{x}\right)^4 \right]\right]\right\}, 
\end{eqnarray}
and (after shifting $z \leftrightarrow 1/y - 1 - z/y$)
\begin{eqnarray}\label{eq:B.2}
R_{{\rm vertex}} (\theta_{{\rm cone}}) &=& \frac{4}{\pi}\, \int^\infty_0\,
\frac{dx}{x^3} \, \int^1_0\,y dy\,\int^\infty_0\,dz\left\{\frac{1}{(z+1)^2}\,
\exp \left[ - \frac{c^2(L) \theta_{{\rm cone}}^2}{(1 + yz)} \, \left( 
\frac{y}{x} \right)^4 \right] \right. \nonumber \\
&-& Re \left. \left[ \frac{1}{\cos^2 (1+i)x\left[ z + \frac{\tan(1+i)x}{(1+i)
x}\right]^2}\,\exp\left[ - \frac{c^2(L)\theta_{{\rm cone}}^2}{y 
\frac{\tan(1+i)x}{(1+i)x} + 1 + yz - y} \left( \frac{y}{x}\right)^4
\right]\right]\right\} . 
\end{eqnarray}

The subtractions in (\ref{eq:B.1}) and (\ref{eq:B.2}), the $\tilde\kappa = 
0$ term in (\ref{eq:2.2}), are taken at fixed $\underline{U}^2$ rather than at 
fixed $\theta_{{\rm cone}}^2$, as is necessary for the subtracted term to 
be medium independent. 


\section{Behaviour for large $\theta_{{\rm cone}}$} 
\label{appendix-c}

From the expressions (\ref{eq:B.1}) and (\ref{eq:B.2}) one may 
evaluate the behaviour of $R(\theta_{{\rm cone}})$ at large values of 
$c(L) \theta_{{\rm cone}} \gg 1$. In view of the 
exponential dependence on $c(L)\theta_{{\rm cone}}$ of the integrands we 
conclude that this behaviour is controlled by the endpoint in the variable
$y$, $y = 0$. Expanding the integrand in (\ref{eq:B.1}) arround this 
point we find 
\begin{eqnarray}\label{eq:C.1}
R_{{\rm on-shell}} (\theta_{{\rm cone}}) & & \mathop{\longrightarrow}\limits_
{c(L)\theta_{{\rm cone}} \gg 1} \, \frac{4}{\pi} \, Re \, \int^\infty_0 \,
\frac{dx}{x^3} \left[ 1 - \left( \frac{1 + i ) x}{\sin(1+i)x} \right)^2 \right]
\nonumber \\
&\times & \int^\infty_0 \, dy \, \int^1_0 dz \exp \left[ - \frac{c^2(L)
\theta^2_{{\rm cone}}}{(1-z)} \, \left( \frac{y}{x} \right)^4 \right] . 
\end{eqnarray}
Performing first the Gaussian $y^2$-integration
and then the $z$-integration we obtain the asymptotic behaviour 
\begin{eqnarray}\label{eq:C.2}
R_{{\rm on-shell}} (\theta_{{\rm cone}}) & & \longrightarrow \frac{4}{5} \,
\frac{\Gamma (1/4)}{\pi} \, \frac{1}{\sqrt{c(L)\theta_{{\rm cone}}}} 
\, Re \, \int^\infty_0 \, \frac{dx}{x^2} \left[ 1 - 
\left( \frac{(1+i)x}{\sin (1+i)x} \right)^2 \right] \nonumber \\
& & = \frac{4}{5} \, \frac{\Gamma (1/4)}{\pi} \, \frac{1}{\sqrt{c(L) 
\theta_{{\rm cone}}}} , 
\end{eqnarray}
The $x$-integration is explicitly given by noting that 
\begin{equation}\label{eq:C.3}
Re \, \int^\infty_0 \, dx \, \frac{d}{dx} \left[ - \frac{1}{x} + 
\frac{(1+i)}{\tan (1+i)x} \right] = 1 . 
\end{equation}
The corresponding expansion around $y \simeq 0$ in (\ref{eq:B.2}) leads to 
\begin{eqnarray}\label{eq:C.4}
R_{{\rm vertex}} (\theta_{{\rm cone}}) & & \mathop{\longrightarrow}\limits_
{c(L) \theta_{{\rm cone}}\gg 1} \, \frac{4}{\pi} \, Re \, \int^\infty_0 \,
\frac{dx}{x^3} \, \int^1_0 \, y dy \, \int^\infty_0\, dz \nonumber \\
&\times&\left[ \frac{1}{(z+1)^2} - \frac{1}{\cos^2 (1 +i) x \left[ z + 
\frac{\tan(1+i)x}{(1+i)x}\right]^2} \right] \exp \left[ - c^2(L)
\theta^2_{{\rm cone}} \left( \frac{y}{x}\right)^4\right] , 
\end{eqnarray}
where the $z$-integration can be performed immediately. The 
$y$-integral leads to
\begin{equation}\label{eq:C.5}
\frac{1}{2} \, \int^1_0 \, dy^2 \, \exp \left[ - \frac{c^2 (L) 
\theta^2_{{\rm cone}}}{x^4} (y^2)^2 \right] = \frac{\sqrt\pi}{4}\,
\frac{x^2}{c(L)\theta_{{\rm cone}}} \, {\rm erf}\, 
\left( \frac{c(L) \theta_{{\rm cone}}}{x^2} \right) , 
\end{equation}
where we keep the erf-function \cite{Abram}
in order to have a finite $x$-integral at 
fixed, but large $c(L)\theta_{{\rm cone}}$,
\begin{eqnarray}\label{eq:C.6}
R_{{\rm vertex}} (\theta_{{\rm cone}}) & & \mathop{\longrightarrow}\limits_
{c(L)\theta_{{\rm cone}} \gg 1} \, \frac{1}{\sqrt\pi} \,
\frac{1}{c(L) \theta_{{\rm cone}}} \nonumber \\
&\times & Re \, \int^\infty_0 \, \frac{dx}{x} \, {\rm erf} \, \left(
\frac{c(L)\theta_{{\rm cone}}}{x^2} \right) \left[ 1 - \frac{(1+i)x}{\sin
(1+i)x \cos(1+i)x} \right] . 
\end{eqnarray}
In Fig.~7
we show that the asymptotic behaviour of (\ref{eq:C.2}) and (\ref{eq:C.6}) is 
also confirmed numerically \cite{Mathem}.
The large $\theta_{{\rm cone}}$-behaviour of $R(\theta_{{\rm cone}})$ 
is dominated by $R_{{\rm on-shell}} (\theta_{{\rm cone}})$ given by 
(\ref{eq:C.2}).

\begin{figure}
\centering
\epsfig{bbllx=35,bblly=210,bburx=510,bbury=600,
file=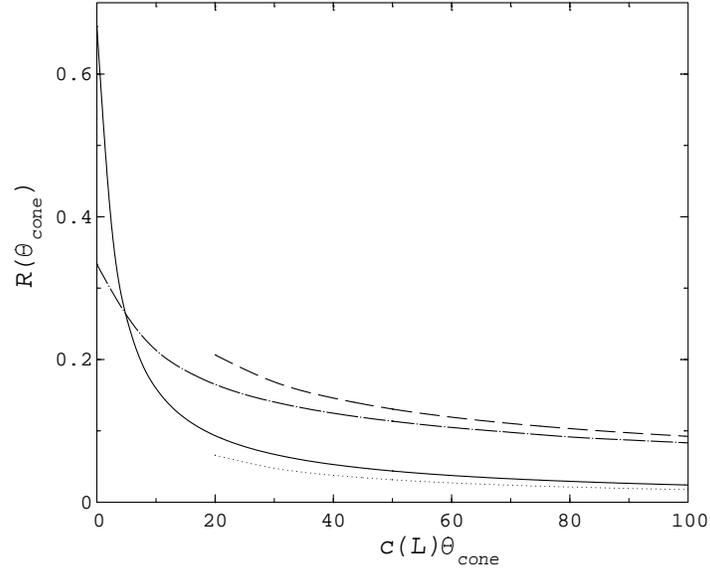,width=96mm,height=80mm}
\caption{\label{fig:graph7}$R_{{\rm on-shell}}$ (dashed-dotted) and 
$R_{{\rm vertex}}(\theta_{{\rm cone}})$ (solid) compared with the 
asymptotic forms (dashed and dotted curves) as given by (\ref{eq:C.2}) and 
(\ref{eq:C.6}), respectively.   }
\end{figure}

\end{appendix}

\end{document}